\documentclass[journal=jacsat,manuscript=article]{achemso}

\usepackage[version=3]{mhchem} 

\usepackage{xcolor}

\author{Frederico B. Sousa}
\affiliation{Departamento de F\'isica, Universidade Federal de Minas Gerais, Belo Horizonte, Minas Gerais 30123-970, Brazil}
\alsoaffiliation{Departamento de F\'isica, Universidade Federal de São Carlos, São Carlos, São Paulo 13565-905, Brazil}

\author{Alessandra Ames}
\affiliation{Departamento de F\'isica, Universidade Federal de São Carlos, São Carlos, São Paulo 13565-905, Brazil}

\author{Mingzu Liu}
\affiliation{Department of Physics, The Pennsylvania State University, University Park, PA 16802, United States of America}
\alsoaffiliation{Center for 2-Dimensional and Layered Materials, The Pennsylvania State University, University Park, PA 16802, United States of America}

\author{Pedro L. Gastelois}
\affiliation{Centro de Desenvolvimento de Tecnologia Nuclear - CDTN, Belo Horizonte, Minas Gerais 31270-90, Brazil}

\author{Vinícius A. Oliveira}
\affiliation{Departamento de F\'isica, Universidade Federal de São Carlos, São Carlos, São Paulo 13565-905, Brazil}

\author{Da Zhou}
\affiliation{Department of Physics, The Pennsylvania State University, University Park, PA 16802, United States of America}
\alsoaffiliation{Center for 2-Dimensional and Layered Materials, The Pennsylvania State University, University Park, PA 16802, United States of America}

\author{Matheus J. S. Matos}
\affiliation{Departamento de F\'isica, Universidade Federal de Ouro Preto, Ouro Preto, Minas Gerais 35400-000, Brazil}

\author{Helio Chacham}
\affiliation{Departamento de F\'isica, Universidade Federal de Minas Gerais, Belo Horizonte, Minas Gerais 30123-970, Brazil}

\author{Mauricio Terrones}
\affiliation{Center for 2-Dimensional and Layered Materials, The Pennsylvania State University, University Park, PA 16802, United States of America}
\alsoaffiliation{Department of Physics, The Pennsylvania State University, University Park, PA 16802, United States of America}
\alsoaffiliation{Department of Materials Science and Engineering, The Pennsylvania State University, University Park, PA 16802, United States of America}

\author{Marcio D. Teodoro}
\affiliation{Departamento de F\'isica, Universidade Federal de São Carlos, São Carlos, São Paulo 13565-905, Brazil}

\author{Leandro M. Malard}
\affiliation{Departamento de F\'isica, Universidade Federal de Minas Gerais, Belo Horizonte, Minas Gerais 30123-970, Brazil}
\email{lmalard@fisica.ufmg.br}

\title[Magneto-PL in Defective WS2 and WSe2]
    {Strong magneto-optical responses of an ensemble of defect-bound excitons in ambient exposed WS$_2$ and WSe$_2$ monolayers}
    
\abbreviations{IR,NMR,UV}

\begin{document}



\begin{abstract}

Transition metal dichalcogenide (TMD) monolayers present a singular coupling in their spin and valley degrees of freedom. 
Moreover, by applying an external magnetic field it is possible to break the energy degeneracy between their K and $-$K valleys. 
This valley Zeeman effect opens the possibility of controlling and distinguishing the spin and valley characters of charge carriers in TMDs by their optical transition energies, making these materials promising for the next generation of spintronic and photonic devices. 
However, the free excitons of pristine TMD monolayers present a moderate valley Zeeman splitting of $\approx 0.23$~meV/T. 
Therefore, alternative excitonic states with higher magnetic responses are mandatory for application purposes. 
Here, we investigate the magneto-optical properties of ambient exposed WS$_2$ and WSe$_2$ monolayers by circularly polarized magneto-photoluminescence experiments at cryogenic temperatures. 
A broad lower energy photoluminescence emission related to an ensemble of defects is observed, presenting remarkable valley-related splittings of $\approx 1.45$~meV/T and $\approx 1.11$~meV/T for WS$_2$ and WSe$_2$ monolayers, respectively.
In addition, we report a significant valley polarization of charge carriers in the defect mid-gap states induced by the external magnetic field. 
We explain this valley-polarized population and enhanced valley-related splitting in terms of imbalanced intervalley relaxations, leading to a magnetic field-dependent distribution of charge carriers in multiple defect levels. 
This effect, together with the individual Zeeman shiftings of the mid-gap states, explains the strong magneto-optical responses observed. 
Our work uncovers the singular potential of manipulating the light emission of ambient exposed TMD monolayers by an external magnetic field.

\end{abstract}

\section{Introduction}

Transition metal dichalcogenides (TMDs) emerged as promising van der Waals semiconductors for the next generation of two-dimensional (2D) technologies due to their singular optical and electronic properties~\cite{manzeli20172d}. 
Similar to conventional semiconductors, the inherent properties of 2D TMDs can be strongly affected by the presence of impurities over the crystal lattice~\cite{Lin2016,Liang2021,sousa2024effects}. 
Hence, there is a great effort to improve the fabrication techniques of these materials to avoid undesirable impurities or to engineer defects to induce novel properties~\cite{Lin2016,Liang2021}. 
Nonetheless, several defects can also be incorporated by 2D TMDs after their fabrication if there is no protection from the environment~\cite{Gao2016} --- such as an encapsulation by layers of hexagonal boron nitride~\cite{iqbal2015high,lee2015highly,ahn2016prevention}. 
Despite the challenge in ascertaining the chemical composition of these non-controlled defects~\cite{Mitterreiter2021}, several authors reported reproducible optical emissions related to these impurities incorporated post-growth in different TMD monolayers. 
In particular, these defects induce a broad lower energy photoluminescence (PL) peak --- often called as L-band --- with a sublinear power dependence at low temperatures~\cite{korn2011low,Tongay2013,He2016,Krustok2017,gordo2018revealing,venanzi2019exciton,verhagen2020towards,greben2020intrinsic,Mitterreiter2021,Xu2021}, which is attributed to the convolution of multiple emissions from mid-gap states introduced by an ensemble of defects. 
However, an investigation of how these defects impact the coupling between the light emission of TMD monolayers and their magnetic properties is still lacking.

TMD monolayers exhibit a coupling in their spin and valley degrees of freedom~\cite{Xiao2012} that can be accessed by a selective carrier excitation in both K and $-$K points by circularly polarized light~\cite{Mak2012,Cao2012}. 
Additionally, by applying an external magnetic field, it is possible to control the valence and conduction bands displacements due to an analogous Zeeman effect~\cite{Li2014,Aivazian2015,Srivastava2015,MacNeill2015}. 
Since the K and $-$K valleys present opposite spins, the band displacement is also contrary in each valley, resulting in an energy splitting of the K and $-$K excitons. 
Hence, the possibility of controlling the spin-valley degree of freedom by an energy threshold makes TMD monolayers promising for spin- and valleytronics~\cite{Ahn2020}. 
Free exciton states in pristine TMD monolayers present a weak valley Zeeman splitting of $\approx 0.23$~meV/T~\cite{Li2014,Aivazian2015,Srivastava2015,MacNeill2015,Stier2016,Plechinger2016}, shedding light on the necessity of engineering these materials to obtain enhanced magneto-optical responses. 
In this sense, different defects such as spin-polarized dopants~\cite{Li2020,Zhou2020,sousa2024giant} and vacancies~\cite{Wang2020} showed the ability to tune the valley Zeeman effect of TMD monolayers in a singular manner.
Besides, previous works studied the Zeeman effect in individual non-magnetic localized states in TMD monolayers, reporting valley splittings of $0.4 - 0.7$~meV/T~\cite{He2015,Koperski2015,Srivastava2015-2,Chakraborty2015,koperski2018orbital}.
However, for ambient exposed samples in which there is a convolution of multiple defect PL peaks, i.e., an L-band emission, their magneto-optical response might also depend on the relaxation dynamics between these states in addition to their individual valley Zeeman splittings, which deserves further investigation. 
Given that sample encapsulation should not be feasible for certain applications, a detailed investigation of this topic can greatly contribute to advancing the understanding of the role of defects on the magneto-optical properties of 2D materials.

Here we report a significant tunability of the PL emission associated with an ensemble of defect-bound excitons in ambient exposed WS$_2$ and WSe$_2$ monolayers by magneto-PL experiments with circularly polarized light detection. 
Power- and temperature-dependent PL experiments showed a broad lower energy peak (L-band) that comprises the convolution of multiple emissions from the mid-gap states. 
Magneto-PL measurements unveiled a high control in the energy and linewidth of the L-band in WS$_2$ and WSe$_2$ monolayers by applying external magnetic fields ranging from $-$9~T to 9~T.
Moreover, the L-band displayed a significant degree of circular polarization induced by the external magnetic field, revealing a valley-polarized charge carrier population in the mid-gap states. 
This valley polarization leads to a magnetic field-dependent distribution of charge carriers in the multiple mid-gap states, which plays a major role in the energy and linewidth tunability of the L-band emission.
Finally, we propose imbalanced intervalley relaxations to explain the magnetic field induced valley-polarized population of charge carriers in the defect levels. 
Our work unveils the strong magnetic response of the L-band emission in ambient exposed WS$_2$ and WSe$_2$ monolayers, underlining the unique spin-valley abilities induced by an ensemble of defects.

\section{Results and discussion}

\begin{figure}[htb!]
 \centering
 \includegraphics[width=1.0\linewidth]{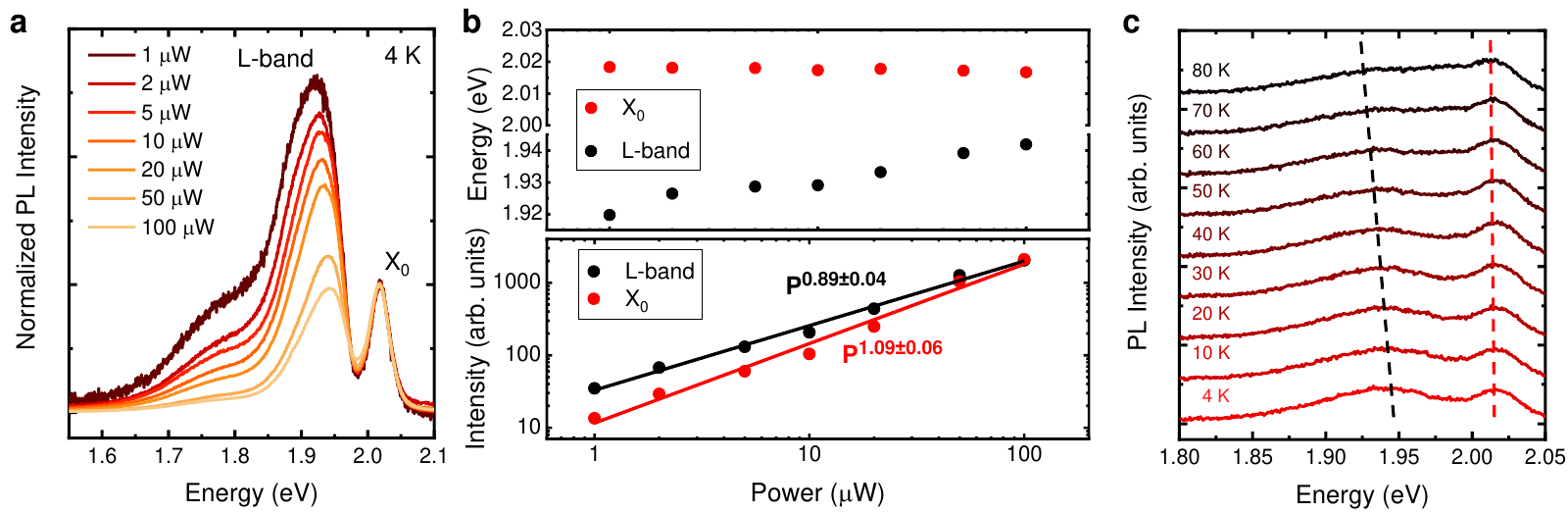} 
 \caption{{\small (a) Power dependent PL spectra at 4~K for a WS$_2$ monolayer showing X$_0$ and L-band emissions. PL spectra are normalized to the intensity of the X$_0$ peak. (b) PL energy (top graph) and intensity (bottom graph) dependencies on the incident power for defect and free exciton emissions. (c) Temperature-dependent PL spectra obtained with a 100~$\mu$W excitation power. All PL spectra were acquired using a 532~nm laser. 
 }}
 \label{fig_WS2_PL}
 \end{figure}

We first performed power-dependent PL experiments to probe the presence of defects in a WS$_2$ monolayer synthesized by chemical vapor deposition (CVD) and study the consequent modifications in its light emission, as shown in Figures~\ref{fig_WS2_PL}a,b. 
The power-dependent PL spectra of Figure~\ref{fig_WS2_PL}a were obtained at 4~K and are normalized to the free exciton (X$_0$) peak ($\sim 2.02$~eV) intensity, also presenting a broad lower energy PL emission. 
While the exciton peak exhibits no relevant shift in its energy and a linear intensity increase with the incident power, the lower energy emission displays a sublinear power dependence and a noticeable energy blueshift, as highlighted in Figure~\ref{fig_WS2_PL}b. 
Moreover, Figure~\ref{fig_WS2_PL}c shows the temperature-dependent PL spectra of the WS$_2$ monolayer, in which the lower energy emission presents an intensity reduction and an energy redshift under increasing temperature. 
Additionally, this lower energy emission vanishes at room temperature, as displayed in Figure~S1. 
As mentioned, a broad lower energy PL emission with sublinear power and temperature dependencies as described above is commonly composed by the convolution of multiple PL peaks from an ensemble of defect states and referred to as L-band. 
In contrast to the optical modifications caused by controlled engineered defects such as substitutional dopants~\cite{Loh2021,sousa2024effects} or spatially localized vacancies~\cite{klein2019site,Mitterreiter2021}, the origin of the L-band is not a consensus~\cite{Mitterreiter2021}. 
Particularly, previous works reported an L-band quenching after annealing treatments~\cite{rogers2018laser,venanzi2019exciton,Mitterreiter2021} and the presence of this broad defect emission only for aged TMD monolayers~\cite{Xu2021}, indicating that these defects might be introduced after ambient exposure.
Besides, although the introduction of these impurities is not controlled, their induced optical properties are reasonably reproducible and agree with our results displayed in Figure~\ref{fig_WS2_PL}~\cite{korn2011low,Tongay2013,He2016,Krustok2017,gordo2018revealing,venanzi2019exciton,verhagen2020towards,greben2020intrinsic,Mitterreiter2021,Xu2021}.
Raman and X-ray photoelectron spectroscopy experiments were also conducted to further characterize the studied samples, as displayed in Figures~S1 and S2, respectively.

\begin{figure}[!htb]
\centering
\includegraphics[width=1.0\linewidth]{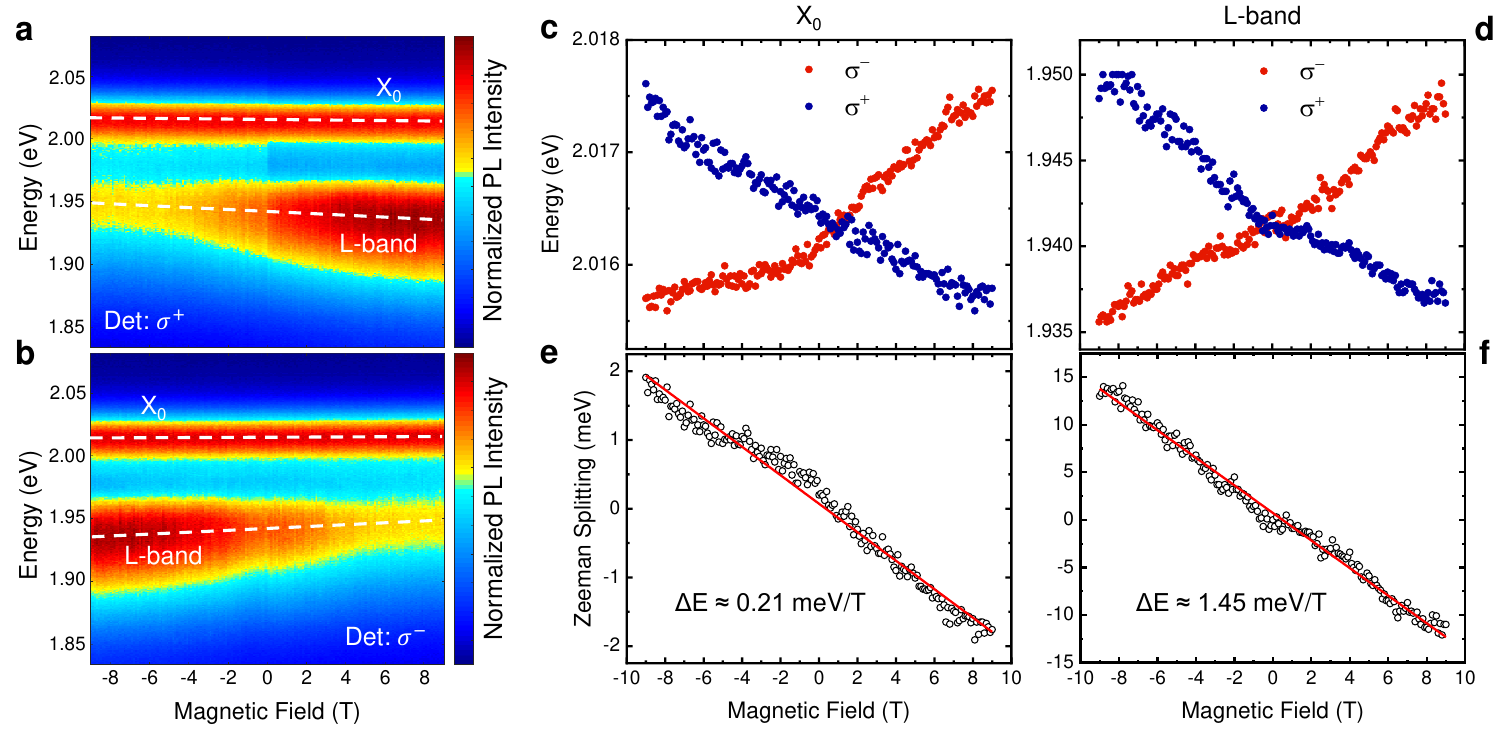} 
\caption{{\small (a,b) Normalized magneto-PL spectra obtained at linearly polarized excitation and $\sigma^+$ (a) and (b) $\sigma^-$  detection with an external magnetic field ranging from $-$9~T to 9~T with steps of 0.1~T. PL spectra were acquired at 4~K, with an excitation power of 100~$\mu$W, and using a 532~nm laser. The energy of X$_0$ (c) and L-band (d) emissions as a function of the magnetic field. Valley splitting of X$_0$ (e) and L-band (f) emissions.}}
\label{fig_WS2_magnetoPL}
\end{figure}

To investigate the circular dichroism of the L-band and its dependence on an external magnetic field, we performed circular polarization magneto-PL experiments at 4~K in the WS$_2$ monolayer. 
The sample was excited by a linearly polarized laser beam with an excitation power of 100~$\mu$W and applying a perpendicular external magnetic field (Faraday geometry) ranging from $-$9~T to 9~T with steps of 0.1~T, while the sample's PL was detected at $\sigma^+$ and $\sigma^-$ circular polarization. 
Figures~\ref{fig_WS2_magnetoPL}a,b exhibit all measured PL spectra acquired at $\sigma^+$ and $\sigma^-$, where it is possible to observe that the L-band experiences a larger influence of the magnetic field on its intensity, linewidth, and position compared to the X$_0$ peak.
This distinct response can also be observed in Figures~\ref{fig_WS2_magnetoPL}c,d, which show the energy position for X$_0$ and L-band emissions at $\sigma^+$ and $\sigma^-$.
We used a Gaussian peak to fit the X$_0$ emission and extract its energy value. 
In contrast, the energy of the L-band emission was determined as the peak maximum position due to its asymmetric character (see Figure~S3 for more details).
From these values, we plotted the difference between the energy of $\sigma^+$ and $\sigma^-$ emissions for X$_0$ and L-band and calculated their respective energy splitting from the data linear fit with respect to the magnetic field, as presented in Figures~\ref{fig_WS2_magnetoPL}e,f. 
A valley Zeeman splitting of $\approx 0.21$~meV/T was obtained for the X$_0$ peak, which is in agreement with previously reported values~\cite{Li2014,MacNeill2015,Srivastava2015,Aivazian2015}. 
Interestingly, a substantial valley-related splitting of $\approx 1.45$~meV/T was revealed for the L-band, following its noticeable energy shift observed in the magneto-PL spectra of Figures~\ref{fig_WS2_magnetoPL}a,b. 
It is worth pointing out that, in contrast to the X$_0$ emission, the magnetic field induced splitting of the L-band cannot be analyzed as a single valley Zeeman splitting.
Since the L-band consists of different emissions associated with optical transitions from several mid-gap states, its magneto-optical response might depend on the combination of the individual Zeeman effect of each defect level and the distinct carrier occupation and dynamics of these states.
Thus, in this work, the term valley-related splitting refers to this complex splitting of the L-band emission instead of the traditional Zeeman splitting of a single defect level.
In addition to the remarkable energy shift, the L-band emission also exhibits a significant linewidth modification with the magnetic field.
As a valley Zeeman effect should mostly affect the emission energy~\cite{Li2014,Aivazian2015,Srivastava2015,MacNeill2015,Stier2016,Plechinger2016}, this spectral shape variation of the L-band evidences the major role of the carrier occupation and dynamics of mid-gap states to the observed phenomenon.
Those mentioned features can also be noted in the magneto-PL line spectra of the WS$_2$ monolayer shown in Figure~S3.
Moreover, we performed magneto-PL experiments in another WS$_2$ monolayer, which confirmed the strong magneto-optical response of the L-band (see Figure~S4).
Finally, Figure~S4 also displays the L-band valley splitting for a magneto-PL measurement carried out with an excitation power of 1~$\mu$W.
At this lower excitation, the energy shift of the L-band presented two distinct slopes for different magnetic field ranges (both showing large valley-related splittings compared to the X$_0$ peak).
This is also consistent with the indication of further contributions beyond the Zeeman shift.

\begin{figure}[!htb]
 \centering
 \includegraphics[width=1.0\linewidth]{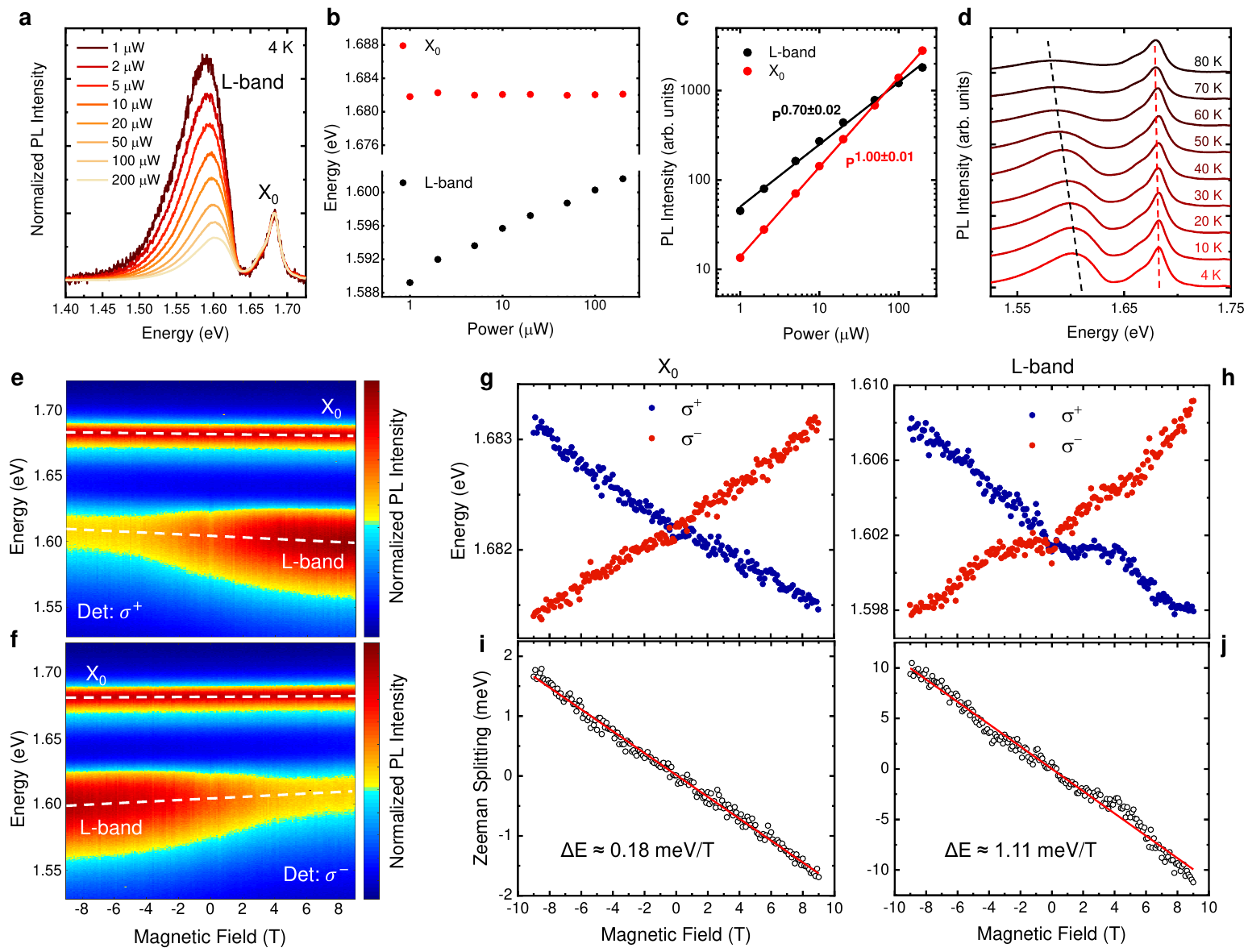} 
 \caption{{\small (a) Power-dependent PL spectra obtained at 4~K for a WSe$_2$ monolayer showing X$_0$ and L-band emissions. PL spectra are normalized to the intensity of the X$_0$ peak. PL energy (b) and intensity (c) power dependencies for X$_0$ and L-band emissions. (d) Temperature-dependent PL spectra obtained with an excitation power of 100~$\mu$W. (e,f) Normalized magneto-PL spectra obtained at 4~K, linearly polarized excitation, and $\sigma^+$ (e) and $\sigma^-$ (f) detection with an external magnetic field ranging from $-$9~T to 9~T with steps of 0.1~T. Energy of X$_0$ (g) and L-band (h) emissions as a functions of the magnetic field. Valley splitting of X$_0$ (i) and L-band (j) emissions. PL spectra were acquired with an excitation power of 100~$\mu$W and using a 660~nm laser.}}
 \label{fig_WSe2_magnetoPL}
 \end{figure}

To further understand the role of defects in the magneto-optical properties of other 2D TMDs, we investigated a CVD-synthesized WSe$_2$ monolayer by the same experiments carried out for the WS$_2$ monolayer. 
Figure~\ref{fig_WSe2_magnetoPL}a shows the power-dependent PL spectra at 4~K for the WSe$_2$ monolayer normalized to the X$_0$ peak intensity and Figures~\ref{fig_WSe2_magnetoPL}b,c display the energy and intensity power dependencies of X$_0$ and L-band emissions. 
Similarly to the WS$_2$ monolayer, an energy blueshift and a sublinear intensity power dependence are noted for the L-band in the WSe$_2$ sample. 
Moreover, temperature-dependent PL measurements shown in Figure~\ref{fig_WSe2_magnetoPL}d reveal another similar behavior compared to the WS$_2$ monolayer, suggesting that both WS$_2$ and WSe$_2$ L-bands have the same origin since they were similarly grown and conserved.

Low-temperature magneto-PL measurements with linearly polarized excitation and circularly polarized detection were also performed for the WSe$_2$ monolayer. 
Figures~\ref{fig_WSe2_magnetoPL}e,f show the PL spectra detected at $\sigma^+$ and $\sigma^-$ and measured by varying the external magnetic fields from $-$9~T to 9~T with steps of 0.1~T. 
PL line spectra obtained at $-$9~T, 0~T, and 9~T are displayed in Figure~S3. 
The energy of X$_0$ and L-band emissions were extracted for all PL spectra (as done for the WS$_2$ data) and are presented in Figures~\ref{fig_WSe2_magnetoPL}g,h. 
In addition to the well-known Zeeman shifting of the X$_0$ peak, it is also noted a relevant shift of the L-band for the WSe$_2$ monolayer. 
The splittings of both emissions were calculated from their energy values and are displayed in Figures~\ref{fig_WSe2_magnetoPL}i,j.
A valley Zeeman splitting of $\approx 0.18$~meV/T was uncovered for the X$_0$ peak, whereas the L-band exhibits a large valley-related splitting of $\approx 1.11$~meV/T.
Furthermore, Figures~\ref{fig_WSe2_magnetoPL}e,f also show the significant dependence of the L-band linewidth on the magnetic field.
Hence, the WSe$_2$ monolayer L-band emission exhibits a similar magneto-optical response compared to the WS$_2$ monolayer.

\begin{figure}[!htb]
 \centering
 \includegraphics[scale=0.6]{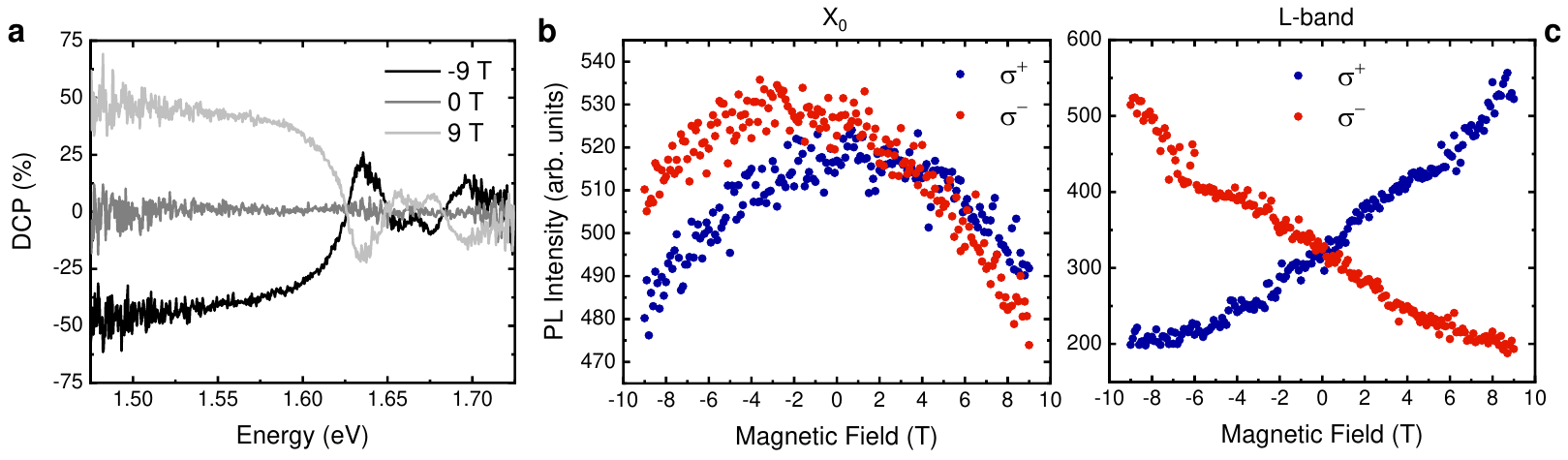} 
 \caption{{\small (a) Degree of circular polarization of a WSe$_2$ monolayer at $-$9~T, 0~T, and 9~T. (b) X$_0$ and (c) L-band PL intensities as a function of the magnetic field.}}
 \label{fig_WSe2_DCP}
 \end{figure}

As observed in Figures~\ref{fig_WS2_magnetoPL} and \ref{fig_WSe2_magnetoPL}, the L-band also presents a strong circularly polarized PL intensity dependence on the external magnetic field in WS$_2$ and WSe$_2$ monolayers. 
To gain deeper insights into this dependence, we analyzed the degree of circular polarization (DCP) of the WSe$_2$ monolayer, given by 
\begin{equation}
    \mathrm{DCP(\%)} = 100\frac{I_{\sigma^+}-I_{\sigma^-}}{I_{\sigma^+}+I_{\sigma^-}},
\end{equation}
in which $I_{\sigma^+}$ and $I_{\sigma^-}$ are the PL intensities at $\sigma^+$ and $\sigma^-$ detections, respectively. 
Figure~\ref{fig_WSe2_DCP}a shows the DCP of the WSe$_2$ monolayer for $-$9~T, 0~T and 9~T. 
They were plotted by subtracting the $\sigma^+$ PL spectrum from the $\sigma^-$ PL spectrum and dividing the result by the sum of $\sigma^+$ and $\sigma^-$ PL spectra.
A negligible DCP can be observed for the whole PL spectral range at 0~T, which is a consequence of the similar PL spectra obtained at $\sigma^+$ and $\sigma^-$ without an external magnetic field.
This absence of DCP is related to the symmetric band structure with anti-symmetric spin moments between K and $-$K valleys at 0~T. 
On the other hand, we can observe a significant variation of the DCP at $-$9~T and 9~T with the emission energy. 
While there is a DCP of $\sim$7.5\% in the X$_0$ spectral range, an increased DCP of $\sim$40\% is observed over the L-band emission. 
The WS$_2$ monolayer also shows a noticeable DCP for the L-band, as shown in Figure~S5.  
In addition to the DCP analysis, we also extracted the intensities of X$_0$ and L-band emissions for all measured magneto-PL spectra in the WSe$_2$ monolayer, as displayed in Figures~\ref{fig_WSe2_DCP}b,c. 
The $\sigma^+$ and $\sigma^-$ X$_0$ peaks exhibit similar PL intensities that present small variations by sweeping the magnetic field.
In contrast, the L-band intensity dependence on the magnetic field is approximately linear, with a positive (negative) slope for the $\sigma^+$ ($\sigma^-$) emission. 
While the L-band emission at 0~T is similar for $\sigma^+$ and $\sigma^-$ detections, the $\sigma^+$ ($\sigma^-$) L-band intensity is $\sim$2.5 times greater than the $\sigma^-$ ($\sigma^+$) L-band intensity at 9~T ($-$9~T), revealing a substantial magnetic field induced valley polarization for the mid-gap states.

Comparing the power-dependent PL spectra (Figures~\ref{fig_WS2_PL}a and \ref{fig_WSe2_magnetoPL}a) with the magnetic field-dependent PL spectra (Figures~\ref{fig_WS2_magnetoPL}a,b and \ref{fig_WSe2_magnetoPL}e,f), it is noted a similar dependence of the L-band on excitation power and magnetic field. 
For power-dependent measurements, the saturation of defect levels leads to modifications in the charge carrier occupation of these mid-gap states, resulting in significant peak energy shift and broadening. 
In this sense, for low excitation powers, the lower energy defect levels are more occupied due to charge carrier relaxation, leading to a lower energy and narrower defect PL emission. 
For high excitation powers, the electronic occupation of the lower energy mid-gap states saturates, resulting in a larger occupation of the higher energy levels and a blueshift and broadening of the defect PL emission.
Figure~S6 schematically represents this light emission power dependence for a direct gap semiconductor with two mid-gap states at low and high excitation power regimes, which can be extrapolated to multiple defect levels to qualitatively describe our scenario.  
Considering all experimental data and discussions presented, we can assume that the carrier occupation also contributes to the magneto-optical responses of the L-band emission, as the Zeeman shifting of each mid-gap level does not completely account for all our observations~\cite{He2015,Koperski2015,Srivastava2015-2,Chakraborty2015,koperski2018orbital}.

\begin{figure}[!htb]
\centering
\includegraphics[scale=0.8]{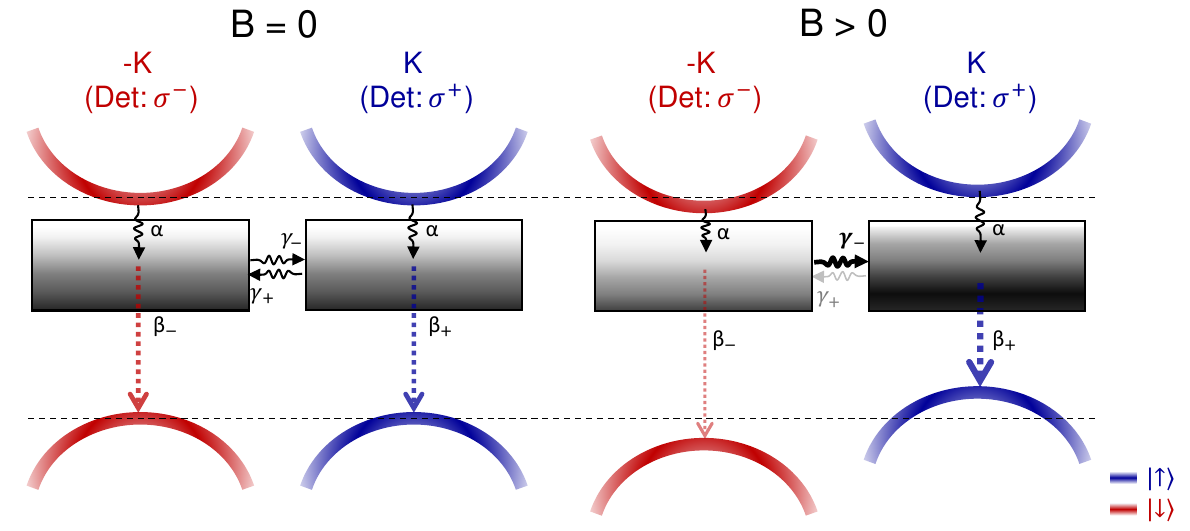} 
\caption{{\small Band structure representation of defective WS$_2$ and WSe$_2$ monolayers for zero and positive magnetic fields.
Degenerate valence, conduction, and continuous defect bands are shown for $\mathrm{B}=0$.
The Zeeman shift of the valence bands and the asymmetric carrier occupation of defect states are reported for $\mathrm{B}>0$.
The blue (red) lines represent spin-up (spin-down) states, while the black rectangles denote the non-spin-polarized continuous defect bands.
The curved arrows depict the relaxation rates ($\gamma_1$ and $\gamma_2$) between K and $-$K mid-gap states.
The dashed arrows describe radiative transitions from the mid-gap states to the valence band. 
The color intensity and thickness of the arrows are related to the transition rate. 
The occupation of the continuum defect bands is proportional to the grayish color scale, in which darker shades of gray indicate a larger carrier population.}}
\label{fig_WS2_spinflip}
\end{figure}

As our magneto-PL experiments were carried out at linearly polarized excitation and constant pumping power (i.e., same excitation at K and $-$K valleys), the magnetic field-dependent valley-polarized carrier occupation of mid-gap states should be induced by relaxation processes after excitation.
Distinct relaxation pathways are reported in the electronic dynamics of TMD monolayers. 
For instance, inter- and intravalley scattering of excited carriers can happen due to electron-phonon and spin-orbit interactions, respectively~\cite{Jiang2021,Robert2021}. 
For mid-gap states, similar relaxations can occur as well.
Particularly, as the existence of localized defect states is associated with the breaking of the crystal translation symmetry, such states shall be more susceptible to intervalley scattering~\cite{linhart2019localized}. 
Additionally, the existence of significant spin-orbit coupling
in both investigated TMD materials might also result in increased spin-flip transition rates. 
Besides, the recombination of electrons in these trapping states is notably slower than of free excitons~\cite{Wang2014,Lagarde2014}, which also favors the relaxation processes in the mid-gap levels.
To better understand the impact of carrier relaxations in the magneto-optical response of the L-band emission, we next present a simple transition rate model for the mid-gap states, following Ref.~\citenum{Aivazian2015}. 
As schematically depicted in Figure~\ref{fig_WS2_spinflip}, this model accounts for a continuum defect band representing the L-band.
Since we pumped our samples with linearly polarized excitation, we consider an equal occupation of the de mid-gap states in K and $-$K valleys from the relaxation of electrons in the conduction band.
From there, the electrons in the defect states can relax between valleys or recombine to the valence band.
Since our reported defects do not involve intentional doping with transition metal atoms, we consider no spin splitting for them.
However, the following analysis would be the same for spin splitted defect levels.
The transition rate equations between K and $-$K defect levels can be written as

\begin{equation}
    \mathrm{\frac{dN_+}{dt} = \alpha I -\gamma_+ N_+ + \gamma_- \mathrm{N_-}} - \beta_+ \mathrm{N_+},
\end{equation}

\begin{equation}
    \mathrm{\frac{dN_-}{dt} = \alpha I -\gamma_- N_- + \gamma_+ \mathrm{N_+}} - \beta_+ \mathrm{N_-},
\end{equation}

\noindent in which $\mathrm{N_+}$ ($\mathrm{N_-}$) is the electronic population at K ($-$K) defect states, $\alpha$ is proportional to the transition rate from the conduction band to the defect bands, I is the laser excitation intensity, $\gamma_+$ ($\gamma_-$) is the transition rate from K ($-$K) to $-$K (K) defect levels, and $\beta_+$ ($\beta_-$) is the decay rate from K ($-$K) defect levels to the valence band.
At 0~T, balanced valley and spin relaxation processes are expected for the mid-gap states.
This means $\gamma_+ = \gamma_-$, $\beta_+ = \beta_-$ and, from Equations~2 and 3, $\mathrm{N_+ = N_-}$, implying no valley or spin polarization induced by carrier relaxation.
In contrast, the carrier scattering within the mid-gap states should present imbalanced intervalley transitions under an external magnetic field~\cite{Slobodeniuk2016,Jiang2021}.
The opposite valley Zeeman shift of the valence bands at K and $-$K points leads to defect-bound excitons with different energies in these two valleys.
Thus, a magnetic field induces an asymmetric population of defect-bound excitons at K and $-$K valleys due to the carrier thermalization to lower energy states.
Particularly, a positive magnetic field causes a greater relaxation to K valley defect states.
In this case, we expect $\gamma_{-} > \gamma_{+}$, resulting in $\mathrm{N_+ > N_-}$ from Equations~2 and 3 and a spin-polarized emission from the mid-gap levels.
Hence, the imbalanced intervalley relaxations for a positive magnetic field lead to an increased population of defect states at K valley that should occupy a large range of lower energy mid-gap levels due to their favorable relaxation.
This assumption is further discussed in a luminescence model related to a constant density of defect states presented in Supporting Section~S1.
This valley-polarized carrier occupation is in agreement with our magneto-PL results, which exhibit enhanced (quenched), broadened (narrowed), and redshifted (blueshifted) $\sigma^+$ ($\sigma^-$) L-band emission for positive magnetic fields.
Following this analysis, the opposite is valid for negative magnetic fields, which also agrees with our experimental results.

Finally, we can correlate the transition rates $\gamma_+$, $\gamma_-$, $\beta_+$, and $\beta_-$ with the measured DCP, since the charge carrier population at K and $-$K defect states can be associated with the circularly polarized L-band emission intensity~\cite{Mak2012,hotger2023spin}. Equations~2 and 3 lead, in a stationary condition, to

\begin{equation}
    \mathrm{N_+} = \frac{\gamma_- + \beta_-/2}{\gamma_+ + \beta_+/2} N_-,
\end{equation}					

\noindent which can be rewritten, with $\mathrm{N_T} \equiv \mathrm{N_+} + \mathrm{N_-}$, $\gamma_1 \equiv \gamma_+ + \beta_+/2$, and $\gamma_2 \equiv \gamma_- + \beta_-/2$, as

\begin{equation}
    \mathrm{N_+} - \mathrm{N_-} = \frac{(\gamma_2-\gamma_1)}{(\gamma_2+\gamma_1)} \mathrm{N_T}.
\end{equation}

Let us now consider that the photoluminescence intensities of the L-band, which involves the recombination of the occupied defect states within the bandgap with holes at the top of the valence band, is mainly determined by the occupation of such defect states, that is, $I_{\sigma^+}/\mathrm{N_+} \approx I_{\sigma^-}/\mathrm{N_-}$.
We therefore obtain, from Equation~5 and the definition of DCP presented in Equation~1, that

\begin{equation}
    \mathrm{DCP(\%)} \approx 100 \frac{(\gamma_2-\gamma_1)}{(\gamma_2+\gamma_1)},
\end{equation}

\noindent where the DCP value is an average value within the energy range of the L-band. 
This allows us to estimate the values of the $\gamma_2/\gamma_1$ ratio from the measured values of DCP as

\begin{equation}
    \frac{\gamma_2}{\gamma_1} \approx \frac{1 + \frac{\mathrm{DCP(\%)}}{100}}{1 - \frac{\mathrm{DCP(\%)}}{100}}.
\end{equation}

From the magneto-PL spectra of Figures~\ref{fig_WSe2_magnetoPL}e,f and using Equation~1, we extracted the DCP of the WSe$_2$ monolayer as a function of the magnetic field, as shown in Figure~S7a.
Finally, from the DCP values and Equation~7, we obtain the estimated values of the $\gamma_2/\gamma_1$ ratio, displayed in Figure~S7b.
It is worth underlining that $\gamma_1$ ($\gamma_2$) represents the total transition rate of electrons from the K ($-$K) valley defect band, which comprises their intervalley relaxation and decay to the valence band.
Hence, we highlight the possibility of obtaining information on the dynamics of an ensemble of defect states by stationary circularly polarized PL experiments.

In summary, we studied WS$_2$ and WSe$_2$ monolayers by power-dependent PL, temperature-dependent PL, and circularly polarized magneto-PL experiments. 
The power- and temperature-dependent PL spectra showed a L-band emission related to an ensemble of defect-bound excitons.  
Additionally, low-temperature circularly polarized PL measurements with a varying external magnetic field were performed to study the magneto-optical response of the L-band. 
A significant energy shift of the convoluted defect emissions corresponding to an intensity enhancement and a spectral broadening was observed for both TMD monolayers under magnetic field variation. 
Substantial valley-related splittings of $\approx 1.45$~meV/T and $\approx 1.11$~meV/T for the L-band were observed in WS$_2$ and WSe$_2$ monolayers, respectively.
In addition, the TMD samples presented a notable valley polarization of charge carriers in the mid-gap states induced by the external magnetic field. 
This valley-polarized population leads to a magnetic field-dependent distribution of charge carriers in the multiple defect levels, which contributes to the large splitting of the L-band.
An imbalanced intervalley transition mechanism was proposed to explain the observed valley polarization.
Our work highlights how an ensemble of incorporated defects can potentialize the tunability of the light emission and the valley polarization of charge carriers in ambient exposed TMD monolayers, shedding light on novel possibilities for spin and valley manipulation in non-encapsulated 2D materials.

\section{Methods}

\subsection{Sample preparation}

WS$_2$ and WSe$_2$ monolayers were prepared using a liquid-phase precursor-assisted deposition method that we developed earlier~\cite{Zhang2020}. In brief, 50 mg of ammonium metatungstate hydrate [(NH$_4$)$_6$H$_2$W$_{12}$O$_{40}$ · xH$_2$O] and 200 mg of sodium cholate hydrate (C$_{24}$H$_39$NaO$_5$ · xH$_2$O) powders were first dissolved in 10 ml of deionized (DI) water to form a precursor solution. The solution was then spin-coated on clean SiO$_2$/Si substrates, which were placed into a quartz tube together with 300 mg of sulfur (S) or 120 mg of selenium (Se) powders upstream. The quartz tube was placed in a two-zone furnace and heated at 800 °C for 15 min with argon (Ar) streaming (for the WSe$_2$ synthesis, an argon/hydrogen (Ar/H$_2$) mixture was applied instead). After the reaction, the whole setup was cooled down to room temperature naturally with Ar protection. PL and Raman spectroscopy experiments were employed to confirm the monolayer thickness of the studied samples, as shown in Figure~S1.

\subsection{Spectroscopy measurements}

The temperature, power, and magnetic field dependent PL spectroscopy experiments were carried out in a confocal microscope coupled to a magneto-cryostat (Attocube --- Attodry 1000). We performed these measurements by varying the temperature from 4~K to 80~K and the applied magnetic field from $-$9~T to 9~T, in which the magnetic field direction was perpendicular to the sample plane. The samples were excited by a linear polarized laser beam with excitation wavelengths of 532~nm (Cobolt --- 08 series) and 660~nm (Toptica --- iBeam) for WS$_2$ and WSe$_2$ monolayers, respectively. The circularly polarized emissions were filtered by a linear polarizer and a quarter-wave plate and detected in a spectrometer equipped with a sensitive CCD camera (Andor --- Shamrock-Idus).

\begin{acknowledgement}

F.B.S., P.L.G., M.J.S.M., H.C., and L.M.M. acknowledge financial support from CNPq, CAPES, FAPEMIG (APQ-01171-21), FINEP, Brazilian Institute of Science and Technology (INCT) in Carbon Nanomaterials and Rede Mineira de Materiais 2D (FAPEMIG). A.M., V.A.O., and M.D.T. acknowledge financial support from FAPESP (2015/13771-0 and 2022/10340-2). M.T., M.L., and D.Z. also thank AFOSR for financial support (FA9550-23-1-0447).

\end{acknowledgement}





\providecommand{\latin}[1]{#1}
\makeatletter
\providecommand{\doi}
  {\begingroup\let\do\@makeother\dospecials
  \catcode`\{=1 \catcode`\}=2 \doi@aux}
\providecommand{\doi@aux}[1]{\endgroup\texttt{#1}}
\makeatother
\providecommand*\mcitethebibliography{\thebibliography}
\csname @ifundefined\endcsname{endmcitethebibliography}
  {\let\endmcitethebibliography\endthebibliography}{}

\clearpage

\setcounter{figure}{0}
\renewcommand{\thefigure}{S\arabic{figure}}

\paragraph*{This Supporting Information includes:\newline}
   %

    $\bullet$ Figure~S1. Room temperature PL spectrum for the WS$_2$ monolayer. \\
    $\bullet$ Figure~S2. XPS experiments for WS$_2$ and WSe$_2$ monolayers. \\
    $\bullet$ Figure~S3. Magneto-PL spectra for WS$_2$ and WSe$_2$ monolayers. \\
    $\bullet$ Figure~S4. Free exciton and L-band valley splittings for the WS$_2$ monolayer. \\
    $\bullet$ Figure~S5. Degree of circular polarization for the WS$_2$ monolayer. \\
    $\bullet$ Figure~S6. Representation of power-dependent light emission from mid-gap states. \\
    $\bullet$ Figure~S7. Degree of circular polarization and transition rates. \\
    $\bullet$ Figure~S8. Luminescence models for a continuum of defect states. \\

\noindent    $\bullet$ Section~S1. Luminescence models for a continuum of defect states \\


\newpage
\begin{figure}[!htb]
	\centering
	\includegraphics[width=1.0\linewidth]{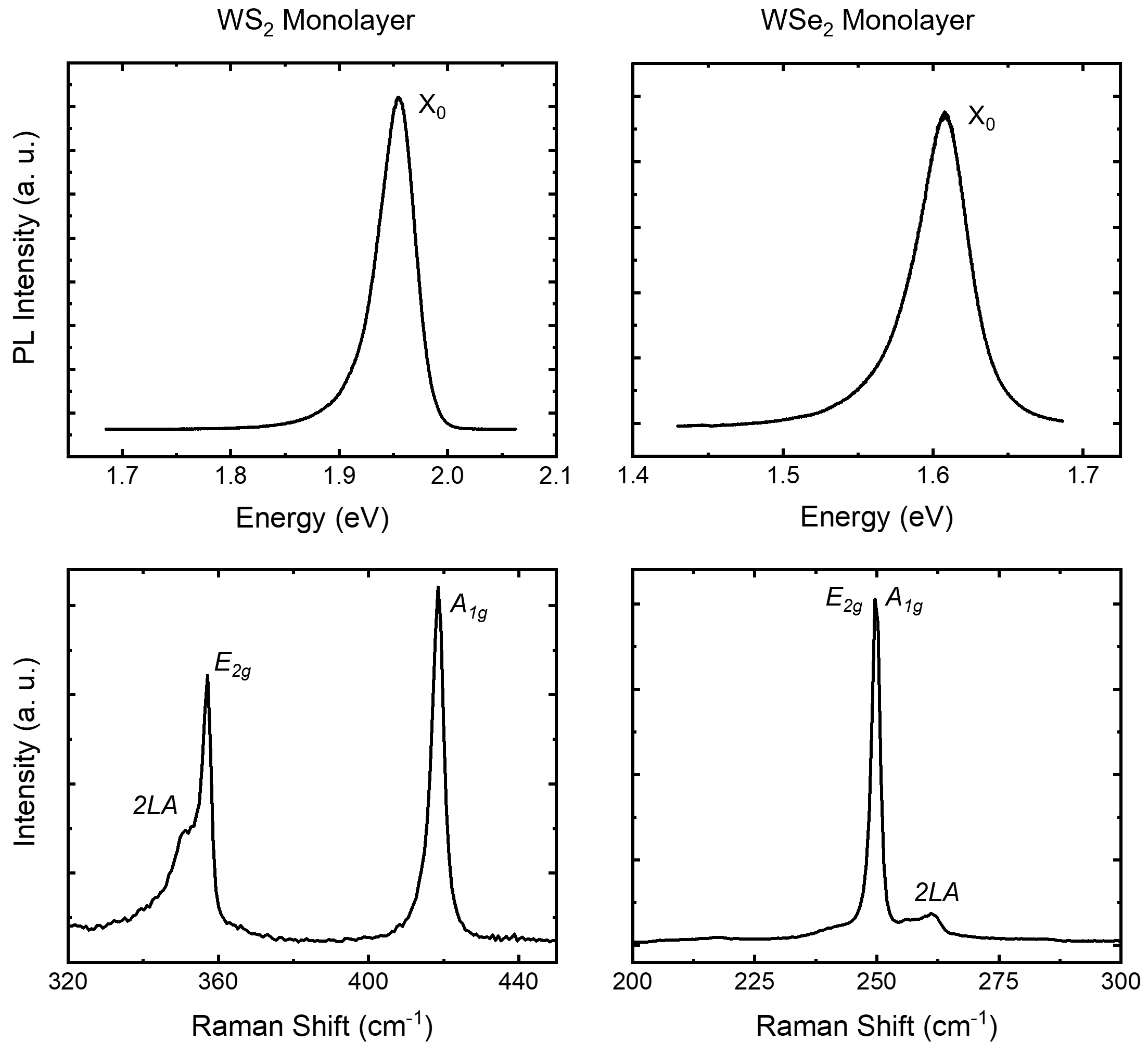} 
	\caption{\small Room temperature PL and Raman spectra for WS$_2$ and WSe$_2$ monolayers. The PL spectrum exhibits no L-band emission at room temperature and a X$_0$ emission at 1.96~eV (1.61~eV) for the WS$_2$ (WSe$_2$) monolayer. The Raman spectrum presents $E_{2g}$, $A_{1g}$, and $2LA$ modes at 357~cm$^{-1}$ (249~cm$^{-1}$), 418~cm$^{-1}$ (250~cm$^{-1}$), and 351~cm$^{-1}$ (261~cm$^{-1}$) for the WS$_2$ (WSe$_2$) monolayer, respectively.}
\end{figure}


\begin{figure}[!htb]
	\centering
	\includegraphics[width=1.0\linewidth]{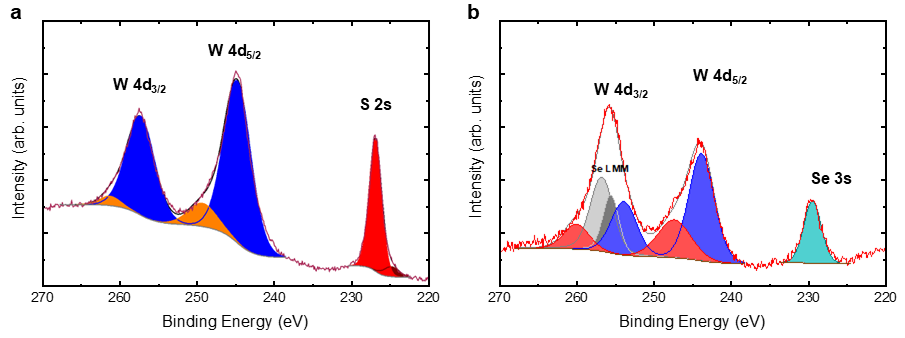} 
	\caption{\small 
    We employed X-ray photoelectron spectroscopy (XPS) measurements to further characterize the studied samples. 
    The valence state of each element present in WS$_2$ and WSe$_2$ monolayers was investigated using the high-resolution XPS spectra analysis. 
    Here, the analysis was used to validate the existence of a minor fraction of defects in our ambient exposed sample. 
    (a,b) XPS spectrum of W~4d, S~2s, and Se~3s core-level peaks for WS$_2$ (a) and WSe$_2$ (b) monolayers.
    While the predominant fitted peaks (in blue) are related to the chemical state doublet of the W-S or W-Se bond, the adjacent fitted peaks indicate the presence of a distinct chemical state doublet in a higher binding energy.}
\end{figure}


\begin{figure}[!htb]
	\centering
	\includegraphics[width=1.0\linewidth]{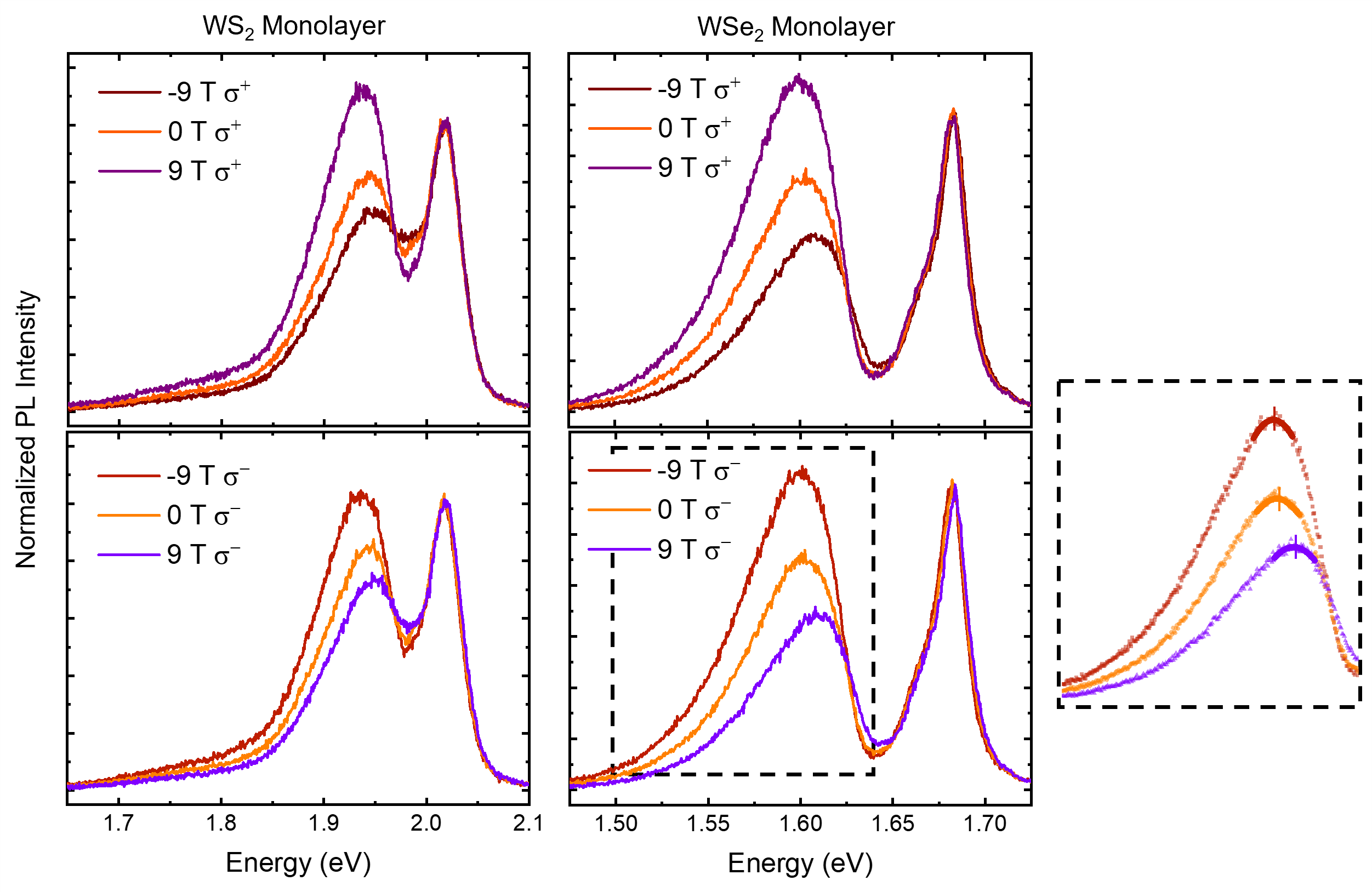} 
	\caption{\small PL spectra of WS$_{2}$ and WSe$_{2}$ monolayers obtained at linearly polarized excitation and $\sigma^+$ and $\sigma^-$ detections with external magnetic fields of $-$9~T, 0~T, and 9~T. The spectra are normalized by the free exciton peak intensity. The inset graph highlighted in the dashed rectangle displays the method used to determine the energies of the L-band emission. In our analysis, we define the L-band energy as the position of the peak maximum. Thus, to obtain this position we fit a narrow spectral region close to the L-band maximum by a Gaussian peak due to the asymmetry of the defect emission, as shown in the inset graph.}
\end{figure}


\begin{figure}[!htb]
	\centering
	\includegraphics[width=1.0\linewidth]{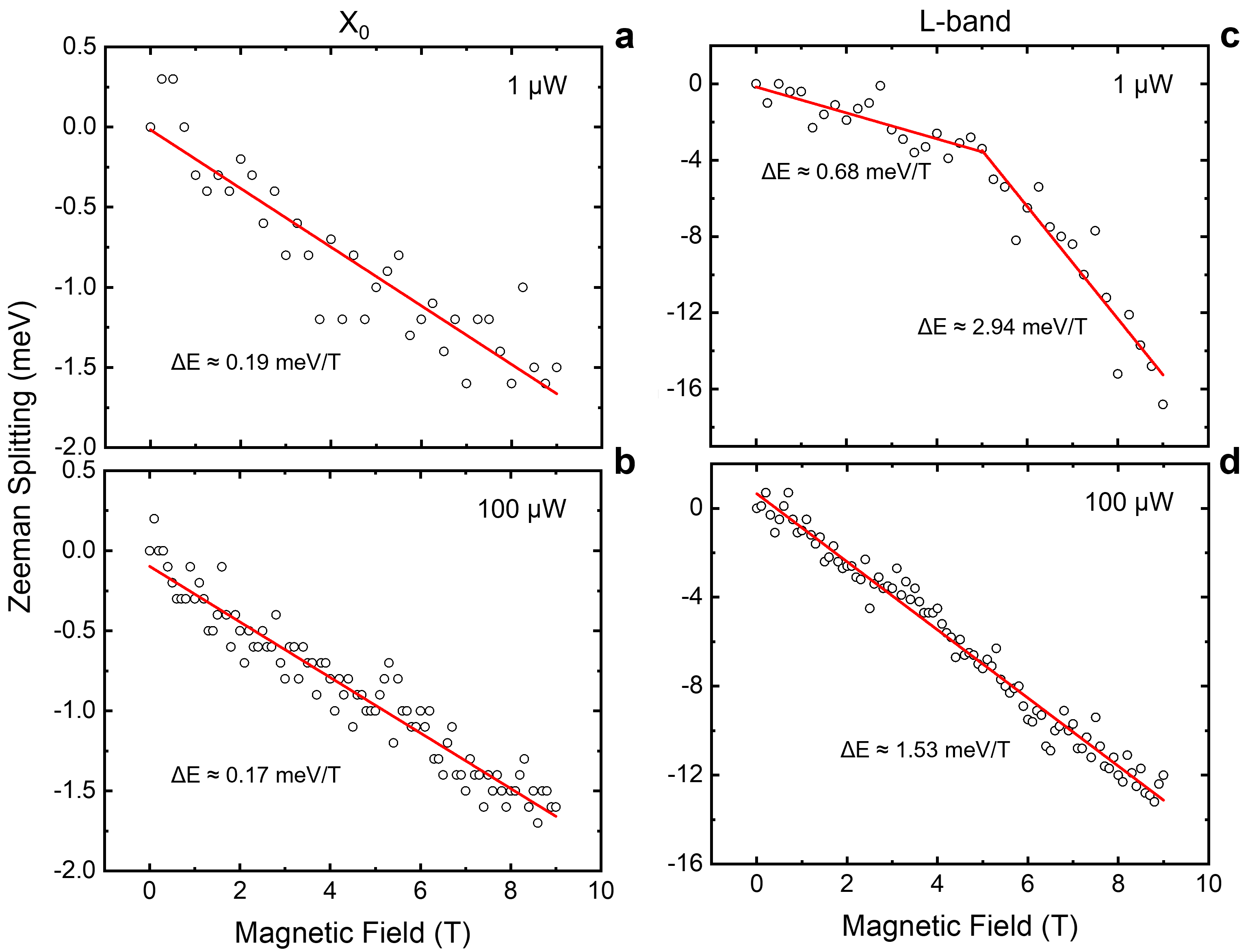} 
	\caption{\small (a)--(d) Valley splittings for X$_0$ (a,b) and L-band (c,d) emissions of a WS$_{2}$ monolayer measured with excitation powers of 1~$\mu$W (a,c) and 100~$\mu$W (b,d). The X$_0$ peak shows a valley Zeeman splitting for both powers. In contrast, while the L-band exhibits an enhanced valley splitting of $\approx 1.53$~meV/T for a 100~$\mu$W excitation (similar to the result presented in the manuscript), we can note that the L-band valley splitting is dependent on the magnetic field under an excitation power of 1~$\mu$W. These results confirm the dependence of the L-band magneto-optical response on the electronic occupation of the mid-gap states. These experiments were performed in a different flake from that presented in Figure~2.}
\end{figure}


\begin{figure}[!htb]
	\centering
	\includegraphics[width=1.0\linewidth]{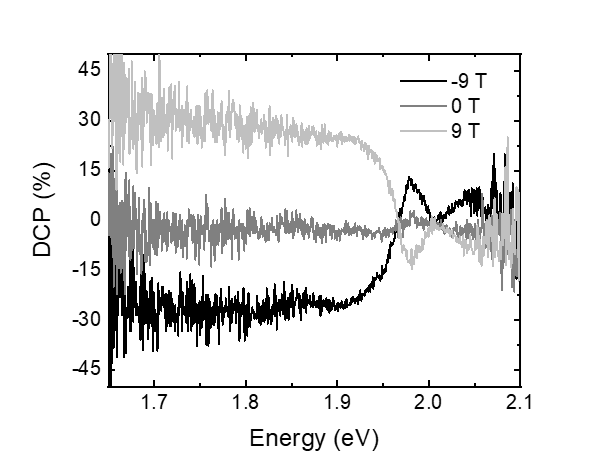} 
	\caption{\small Degree of circular polarization of the WS$_{2}$ monolayer for $-$9~T, 0~T and 9~T calculated from the $\sigma^+$ and $\sigma^-$ PL spectra shown in Figure~2.}
\end{figure}


\begin{figure}[!htb]
 \centering
 \includegraphics[width=1.0\linewidth]{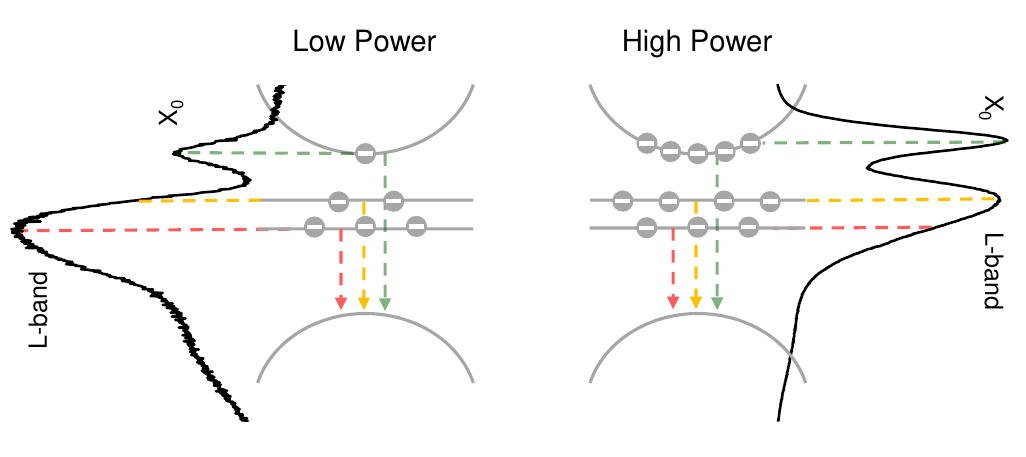} 
 \caption{{\small Illustrative representation of the radiative recombination from the conduction band and mid-gap states for a low and high incident power regime in a direct gap semiconductor. The PL spectra of the WS$_2$ monolayer obtained with excitation powers of 1~$\mu$W and 100~$\mu$W are displayed beside each band structure diagram to associate their energy shift with the electronic occupation of the mid-gap states.}}
 \label{fig_WS2_power}
\end{figure}


\begin{figure}[!htb]
	\centering
	\includegraphics[width=1.0\linewidth]{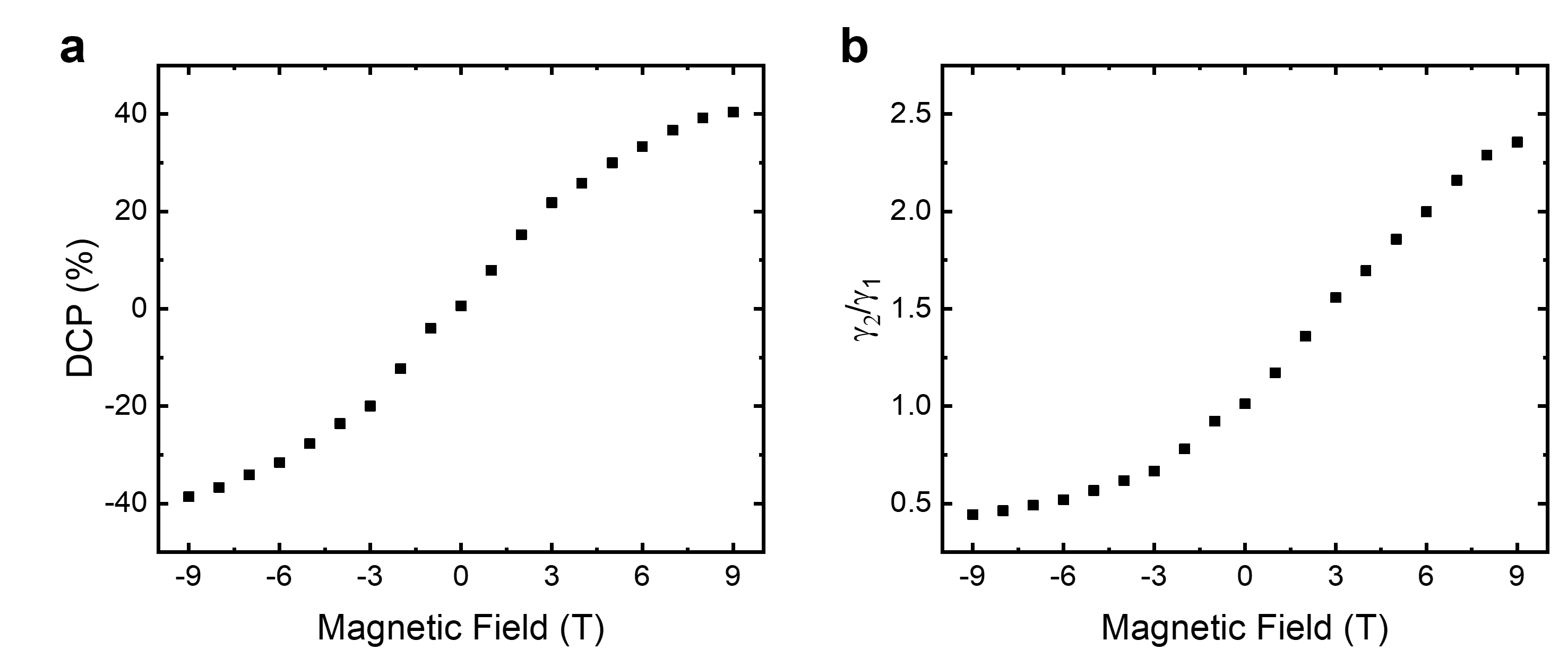} 
	\caption{\small (a) Magnetic field-dependent degree of circular polarization of the WSe$_{2}$ monolayer within the energy range of the L-band calculated from the $\sigma^+$ and $\sigma^-$ PL spectra shown in Figure~3. (b) $\gamma_2/\gamma_1$ ratio calculated from Equation~7 and the DCP values shown in (a).}
\end{figure}


\clearpage
\section*{\large{%
Section~S1. Luminescence models for a continuum of defect states
}}

Let us analyze the consequences of Eqs. (2)-(9) to the PL peak positions as a function of the magnetic field in our studied samples where many types of defects exist. 
We will consider, for simplicity, a continuum of such types of defects, and that these defects are distributed homogeneously in energy inside the bandgap region. 
This leads, as shown in Figure~S7, to a defect band within the bandgap, consisting of a constant density of defect states, which we will consider initially unoccupied, between energies $\epsilon_1$ and $\epsilon_2$. 
We will also consider two alternative models (a) and (b) for the luminescence dynamics, shown schematically in Figure~S7(a) and S7(b), respectively. 
In model (a), we will consider that the photoexcited electrons at the conduction band decay, in a first step, to states at the top of the defect band, and then, in a second step, these occupied defect states recombine with holes at the top of the valence band at $\epsilon_v$, emitting photons with average photon energy $E$. 
A plausible reasoning behind the occupation of defect states in this model is that the transition rate from the conduction band states to defect states is inversely proportional to their energy difference. 
In model (b), nearly the same occurs as in (a), with the difference that the conduction band electrons decay, in the first step, to states at the bottom of the defect band. 
A plausible reasoning behind the occupation of defect states in model (b) is the possibility of thermalization between energy states of neighboring defects, leading to decay of occupation towards the bottom of the defect band.
In model (a), Eqs.~(7) and (8), and Figure~S7, lead to the difference $\Delta E_a = E_+ - E_-$ between the average luminescence photon energies $E_+$ and $E_-$ in the K and $–$K valleys, respectively, as

\begin{equation}
    \Delta E_a = E_+ - E_- = -\frac{1}{2D} (\mathrm{N_+} - \mathrm{N_-}) = -\frac{\mathrm{N_T}}{2D} \frac{(\gamma_2-\gamma_1)}{(\gamma_2+\gamma_1)} = -\frac{\mathrm{N_T}}{2D} \frac{\mathrm{DCP(\%)}}{100},
\end{equation}

\noindent where D is the density of states value at the defect band. 
Likewise, for model (b) we obtain

\begin{equation}
    \Delta E_b = E_+ - E_- = \frac{1}{2D} (\mathrm{N_+} - \mathrm{N_-}) = \frac{\mathrm{N_T}}{2D} \frac{(\gamma_2-\gamma_1)}{(\gamma_2+\gamma_1)} = \frac{\mathrm{N_T}}{2D} \frac{\mathrm{DCP(\%)}}{100}.
\end{equation}

Therefore, both models are consistent with the observed splitting between the L-band energies in the K and $–$K valleys under a magnetic field as a result of carrier dynamics, even without the existence of a “physical” valley Zeeman effect.


\begin{figure}[!htb]
	\centering
	\includegraphics[width=1.0\linewidth]{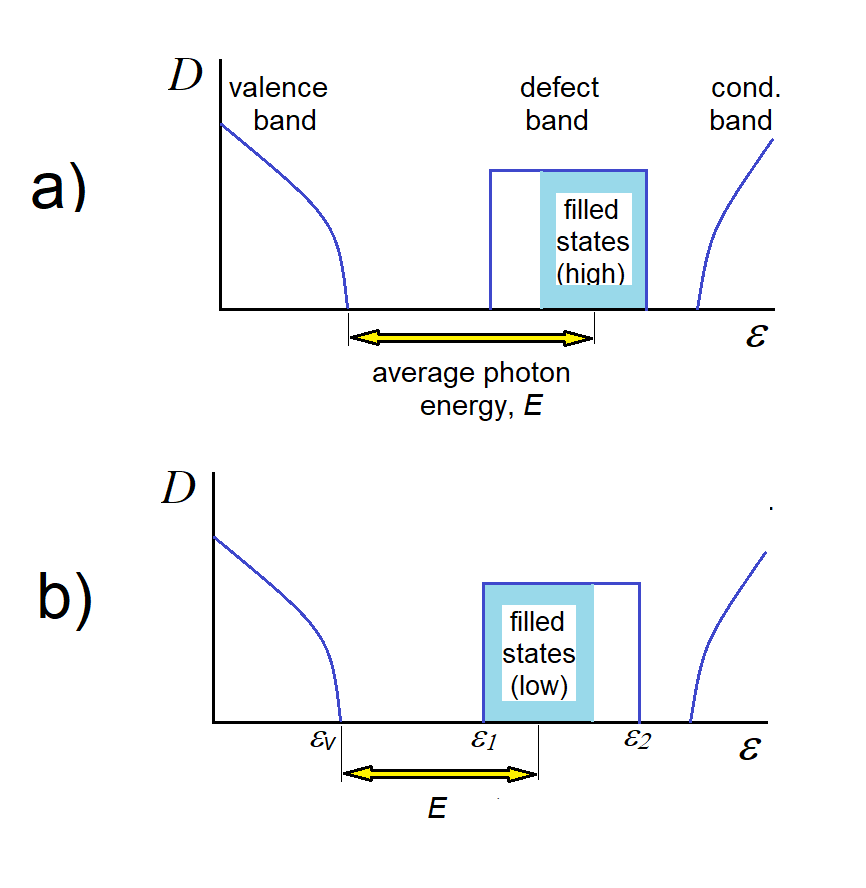} 
	\caption{\small In both models (a) and (b), we assume the existence of a defect band within the bandgap, consisting of a constant density of defect states D, initially unoccupied, between energies $\epsilon_1$ and $\epsilon_2$. In model (a), we consider that photoexcited electrons at the conduction band decay, in the first step, to states at the top of the defect band, and then, in the second step, these occupied defect states recombine with holes at the top of the valence band at $\epsilon_v$, emitting photons with average photon energy E. In model (b), the conduction band electrons decay, in the first step, to states at the bottom of the defect band.}
\end{figure}





\end{document}